\documentclass[a4paper]{article}

\usepackage{INTERSPEECH2020}

\title{Neural Speech Separation Using Spatially Distributed Microphones}
\name{Dongmei Wang, Zhuo Chen, Takuya Yoshioka}
\address{Microsoft, One Microsoft Way, Redmond, WA, USA}
\email{\{dowan, zhuc, tayoshio\}@microsoft.com}

\begin{document}

\maketitle
\begin{abstract}
This paper proposes a neural network based speech separation method using spatially distributed microphones. Unlike with traditional microphone array settings, neither the number of microphones nor their spatial arrangement is known in advance, which hinders the use of conventional multi-channel speech separation neural networks based on fixed size input. 
To overcome this, 
a novel network architecture is proposed that interleaves inter-channel processing layers and temporal processing layers. 
The inter-channel processing layers apply a self-attention mechanism along the channel dimension to exploit the information obtained with a varying number of microphones. 
The temporal processing layers are based on a bidirectional long short term memory (BLSTM) model and applied to each channel independently. 
The proposed network leverages information across time and space 
by stacking these two kinds of layers alternately. 
Our network estimates time-frequency (TF) masks for each speaker, 
which are then used to generate enhanced speech signals either with TF masking or beamforming. 
Speech recognition experimental results show that the proposed method significantly outperforms baseline multi-channel speech separation systems. 
\end{abstract}

\section{Introduction}
\label{sec:intro}
Speech separation research has made a tremendous progress over the past five years thanks to neural network approaches. 
The separation technology is indispensable for dealing with overlapped utterances in automatic speech recognition (ASR) because 
standard ASR frameworks can handle only one speaker at a time. 
Most existing successful speech separation front-ends for far-field ASR use a neural network that takes input from a fixed geometry microphone array \cite{ty_icassp_2018,Bahmaninezhad2019}, where permutation invariant training (PIT)~\cite{dyu_taslp_2017} is often applied to generate time-frequency (TF) masks for each speech source. 
To avoid the spectral distortion caused by TF masking, a beamformer can be created based on the estimated TF masks and applied to the microphone signals~\cite{6516079,HEYMANN2017374,erdogan_interspeech_2016}. 
This type of multi-channel neural speech separation approach has been shown to yield significant ASR performance gains in real meetings~\cite{ty_interspeech_2018,PrincetonASRU2019},

However, the speech separation technology has yet to be mature enough for ad hoc array recordings, where an unknown number of microphones are randomly distributed in a room. 
Ad hoc array processing allows people to use their own devices, e.g., cellphones or laptops, to virtually form a microphone array or enhance a microphone array installed in a room, thus providing greater user flexibility than fix array processing.
It also enables speech processing algorithms to make use of diverse spatial information. With the cloud computing technology, the audio signals from the individual devices can be transmitted to and processed in the cloud for speech separation and transcription. 

Two challenges have to be addressed to utilize spatially distributed microphones.
Firstly, the number and spatial arrangement of the microphones are unknown. Secondly, the individual microphone signals are asynchronous. The latter problem can be largely alleviated by existing methods and is thus out of the scope of this paper~\cite{zliu_iwaenc_2008,Araki18-ICASSP,Yoshioka19c}. 
On the other hand, the first issue hinders the use of 
conventional multi-channel speech separation networks that capitalize on time differences of arrival (TDOAs) obtained from a fixed geometry array. Transform-average-concatenate (TAC) method was proposed very recently~\cite{yiluo_icassp_2020} to address the distributed microphone challenge for an end-to-end speech separation framework called FaSNet~\cite{9003849}. 
However, little has been studied for the TF mask-based approach, which has been more successful on real data (see e.g., \cite{PrincetonASRU2019}) for the fixed array scenarios.


\begin{figure*}[t]
\centering
\includegraphics[scale=0.46]{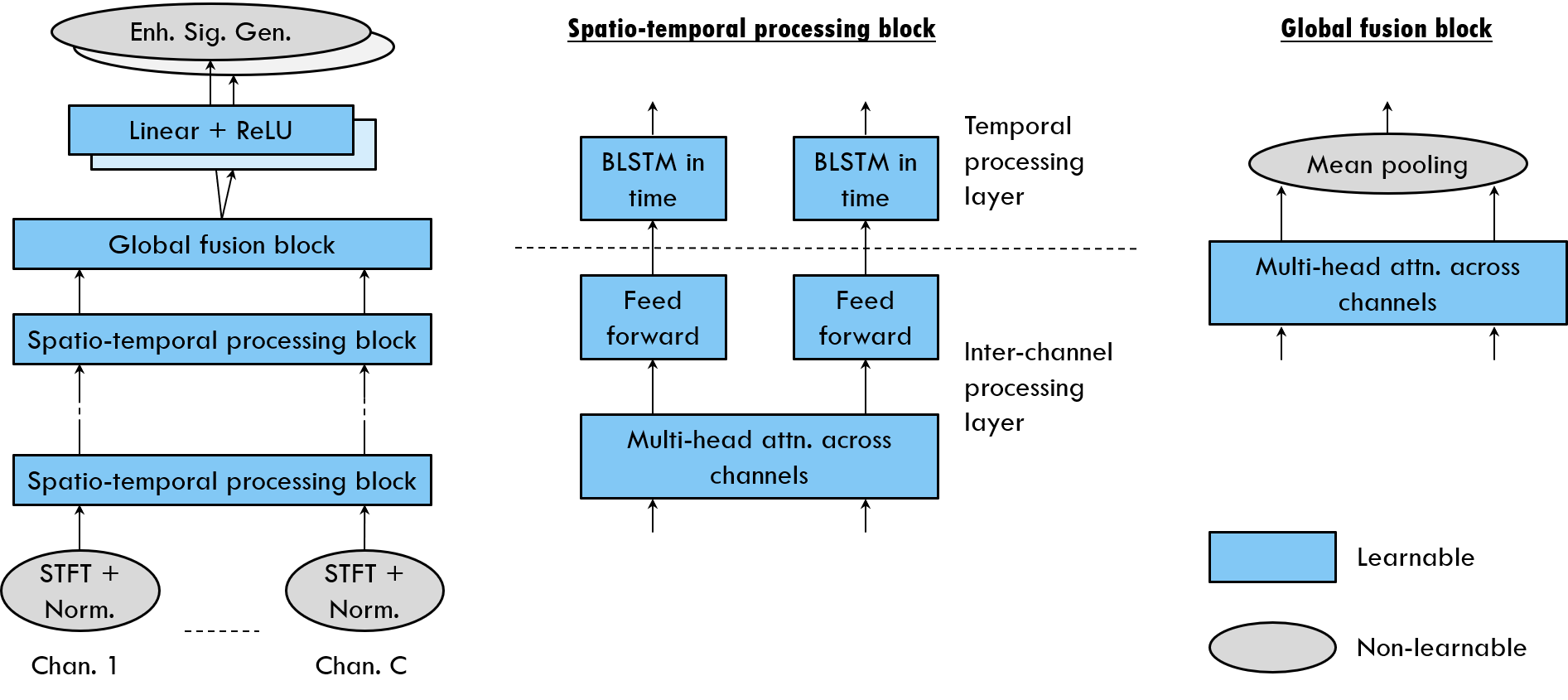}
\caption{Block diagram of proposed model.}
\label{fig: diagram}
\end{figure*}

In this paper, we propose a deep neural network-based speech separation method that is invariant to the number and order of input channels as well as their spatial arrangement. 
The proposed network stacks a set of spatio-temporal processing blocks, each consisting of 
an inter-channel processing layer and a time recurrent BLSTM layer. 
The inter-channel processing layer applies a self-attention mechanism~\cite{transformer_2017} along the channel dimension to leverage spatial diversity in a way that is invariant to the number and order of the input channels. 
Then, the BLSTM layer is applied to each output of the inter-channel processing layer for temporal modeling. By stacking the spatio-temporal processing blocks, 
the proposed method attempts to capture both the inter-channel and temporal correlations from the multi-channel input. 
Finally, global mean pooling layer, which is also invariant to the channel count and order, is applied to fuse the information from all the channels for subsequent mask estimation, which is used to obtain final enhanced speech signals, by either with TF masking or minimum variance distortionless response (MVDR) beamforming.  ASR experiments are performed to show the efficacy of the proposed model on real recordings obtained with multiple cell phones.



\section{Proposed system}
\label{sec:proposed system}

Figure \ref{fig: diagram} shows the processing flow of our proposed speech separation model using spatially distributed microphones. 
%
The proposed model consists of three elements, spatio-temporal processing, global fusion, and mask estimation. After being processed with short time Fourier transform (STFT) and mean and variance normalization, the multi-channel signals firstly undergo multiple spatio-temporal processing blocks, with which the number of channels remain unchanged. Then, a global fusion layer consolidates the multi-channel information. Finally, TF masks are estimated for each speech source by the mask estimation layers.



\subsection{Spatio-temporal processing block}

The spatio-temporal processing block consists of an 
inter-channel processing layer and a temporal processing layer. 
By stacking multiple blocks, or interleaving the inter-channel and temporal processing layers, 
the entire network exploits the information across space and time.

\subsubsection{Inter-channel processing layer}
\label{ssec:cia}

The inter-channel processing layer accepts a 3d tensor of shape (channel, time, feature) as input and 
produces another 3d tensor as output. 
By applying the self-attention mechanism along the channel dimension, the representations of each channel are combined with those of the other channels. 
A multi-head dot-product attention scheme~\cite{transformer_2017}
is used for the combination.


Let $\mathbf{X} = [\mathbf{X}_0, \cdots, \mathbf{X}_{T-1}]$ denote the input
tensor, where $\mathbf{X}_t \in \mathbb{R}^{N\times C}$ is the 
input feature matrix for time $t$ while $T$, $C$, and $N$ represent the numbers of time frames, the input channels, and the features per channel, respectively. 
Let us also denote the number of attention heads by $D$. 
For each attention head, the input features are transformed into query (Q), key (K), and value (V) embedding subspaces of dimension $E$ as follows: 
\begin{align} 
\mathbf{Q}_t^i &= \mathbf{W}_Q^i \mathbf{X}_{t} + \mathbf{b}_Q^i \\
\mathbf{K}_t^i &= \mathbf{W}_K^i \mathbf{X}_{t} + \mathbf{b}_K^i \\
\mathbf{V}_t^i &= \mathbf{W}_V^i  \mathbf{X}_{t} + \mathbf{b}_V^i. 
\end{align}
Matrices $\mathbf{Q}_t^i$, $\mathbf{K}_t^i$, and $\mathbf{V}_t^i$ (all of them in $\mathbb{R}^{E\times C}$) denote the query, key, and value matrices, respectively, for the $i$th attention head at time $t$, which are obtained by the transformations defined by 
$\mathbf{W}_s^i\in \mathbb{R}^{E \times N}$ and $\mathbf{b}_s^i \in \mathbb{R}^{E}$ ($s \in \{K, Q, V\}$). 
Within each head, a cross-channel similarity matrix is computed by taking the product of the query and key matrices. A softmax function is applied to each column of this matrix to obtain an attention matrix $\mathbf{A}_t^i \in \mathbb{R}^{C \times C}$ as 
\begin{equation} 
\mathbf{A}_t^i = \text{softmax} \left( \left( \mathbf{Q}_t^i \right) ^\top \cdot \mathbf{K}_t^i\right). 
\end{equation}
The value matrix $\mathbf{V}_t^i$ is multiplied by the attention matrix as  \begin{equation} 
\mathbf{Y}_t^i = \mathbf{V}_t^i\cdot {\left(\mathbf{A}_t^i\right)}^\top, 
\end{equation}
yielding an intermediate output matrix $\mathbf{Y}_t^i \in \mathbb{R}^{C \times E}$, which is concatenated across the subspaces as 
\begin{equation} 
\mathbf{Y}_t = 
\begin{bmatrix}
\mathbf{Y}_t^0\\
\vdots \\
\mathbf{Y}_t^{D-1} 
\end{bmatrix}
\in \mathbb{R}^{(E\times D) \times C  }. 
\end{equation}
A position-wise single-layer fully connected network with ReLU activation is applied to each channel slice of $\mathbf{Y}_t$ to generate an output, $\mathbf{Z}_t \in \mathbb{R}^{N \times C}$, of the inter-channel processing layer. 
A residual connection is applied between the input and output of the inter-channel processing layer 
to mitigate the gradient vanishing problem.



\subsubsection{Temporal processing layer}
\label{ssec:tmpl_model}

The temporal processing layer applies a shared neural network to each output channel from the inter-channel processing layer to capture the 
temporal correlation within each channel and thereby leverage the dynamics inherent in speech signals.  
Various temporal models may be used, including recurrent layers, time convolutional layers, and self-attention layers with positional encoding. 
In this paper, we choose to use BLSTM with output projection~\cite{Sak14}, whose parameters are shared across all the channels. 


\subsection{Global fusion}
\label{ssec:global_fusion}

After multiple layers of spatio-temporal processing, the number of the channels in the neural network remains the same as the microphone number while the information has been transmitted across the channels and time frames. 
A global fusion layer is introduced to consolidate the information from all the channels in a way that is independent of the number of the input channels. 
Cross-channel multi-head attention followed by 
mean pooling is used in this paper for global fusion.


\subsection{Mask estimation and training criterion}
\label{ssec:mask_esti}

We adopt the TF masking approach~\cite{mask_2013} for speech separation during model training. Two different fully connected feed-forward layers are appended on top of the mean-pooling layer to estimate the TF masks for two target speakers. The estimated masks are applied to the mixed speech spectrum of the first input channel to compute a separated speech spectrum for each source which is processed by inverse short time Fourier transform to generate final time domain separation result.


The entire model is trained with an utterance-level scale invariant signal-to-noise ratio (SISNR) between the separated and clean waveforms~\cite{roux2018sdr}. 
PIT is employed to deal with the output permutation indeterminacy problem.

\subsection{Signal enhancement at test time}
\label{ssec:mvdr}

At test time, there are two things that have to be considered. 
\begin{itemize}
    \item The nonlinear distortion resulting from the TF masking is detrimental to ASR, which can be mitigated by a mask-based MVDR beamforming technique~\cite{7404828}. 
    \item The mask-based MVDR recovers a source image that would have been observed at a selected spatial location, which is often chosen from the microphone positions. In the distributed microphone set-up, some microphones may be much closer to the speaker to be extracted than others. Therefore, for each speaker, we want to pick the microphone that has the highest signal-to-distortion ratio (SDR) for that speaker's signal. 
\end{itemize}

The MVDR beamforming coefficients are estimated for each target speaker by following \cite{erdogan_interspeech_2016}. 
The reference microphone is selected to maximize a posterior SNR~ \cite{erdogan_interspeech_2016}. 
One drawback of using MVDR beamforming for speech separation is that it maintains a unit gain toward a certain direction. 
Therefore, even if the separation mask values are zeros over all TF bins, MVDR cannot completely filter out the interfering speech especially under reverberant conditions. 
This results in a significant increase in insertion errors when the utterances are only partially overlapped, which is usually the case in practice. 
Following \cite{ty_icassp_2018}, 
we resolve this problem by applying VAD masks to the MVDR output to remove the leaked interference. 
The VAD masks are estimated with an initial ASR pass on
the TF-masked signals.

\section{Experiments}
\label{sec:exprmt}

\subsection{Data}
\label{ssec:data}
 
Our training data set is created with simulation from 
the 460-h clean subset of the LibriSpeech corpus~\cite{panayotov2015librispeech}. 
Room impulse responses (RIRs) are generated with the image method
to simulate room reverberation~\cite{image_method_1979, gpu_rir}.
Specifically, a conference-room distributed microphone scenario is simulated as follows. 
i) Randomly choose a room size, a table size, and a reflection coefficient. 
ii) Place the bounding box defined by the table at a random place in the room. 
iii) Randomly choose 10 microphone locations within the bounding box. 
iv) Randomly choose 10 loudspeaker locations around the bounding box. 
v) Run the image method.

A total of 230,000 RIRs are generated offline. 
For each training example, we randomly choose one room and pick up seven microphones and two speaker positions to obtain the RIRs. Two randomly chosen LibriSpeech utterances are convolved with the RIRs to generate two seven-channel reverberant signals. Then, they are overlapped with each other at a randomly determined overlap ratio to generate a seven-channel mixed speech signal. The overlap ratio is sampled from a uniform distribution in $\left[0, 1\right]$. A point source noise is added to each mixed speech signal at an SNR level around 15 dB. 
A validation set is also created in the same way by using different RIRs than those used for the training set.

For evaluation, we created a real recording dataset by using distributed microphones. The recording took place in a conference room (10m $\times$ 5m) with constant air conditioning noise. We used five loudspeakers to continuously playback concatenated LibriSpeech clean test utterances to simulate a meeting scenario with five attendees. The maximum number of simultaneously active speakers was set to two. We split the test set into eight groups. Different overlap ratios were used for different groups, ranging from 5\% to 30\%.
Each group's audio was approximately 10 min long. 
Four cellphones (two iPhones and two Android phones) and three laptops were used as recording devices, which were randomly placed on a table with at least one device in front of each loudspeaker. The audio signals were sampled at 16 kHz. Automatic gain control (AGC) was turned off for all the devices during the recording. 
To facilitate channel synchronization and utterance segmentation, we added  a pure tone signal at the beginning of the playback audio file of each group. 

In this paper, we use the reference tone signal to synchronize the
individual device signals. Also, each mixed utterance was extracted by using the correct segmentation information. 
Evaluation in a more realistic continuous speech separation setting~\cite{chen2020continuous} is out of our current scope. 

\subsection{Separation model training}
\label{ssec:training}

Our speech separation model is configured as follows. 
The input waveform of each channel is transformed into an STFT representation with 257 frequency bins every 16 ms.
Layer normalization is performed on the input spectrum magnitude vectors. 
Three spatio-temporal processing blocks are stacked.
All self-attention layers have 128-dimensional embedding spaces and 
eight attention heads. 
A position-wise fully connected layer transforms this onto a 257-dimensional space. The BLSTM layers have 512 cells for each direction, followed by a projection layer that maps onto a 257-dimensional space. 
In each spatio-temporal block, a residual connection is added. 
A global fusion layer is appended on top of the spatio-temporal processing blocks, which is followed by two 257-dimensional mask estimation layers, one for each speech source.


During training, a learning rate is decayed by half when the SISNR measured on the validation set does not decrease for three consecutive epochs.

\subsection{Comparison systems}

Two baseline systems are built to examine the effectiveness of the proposed method. 
One system, referred to as a multi-stream approach, applies a monaural separation system to each microphone signal and then consolidates the separation results from all the channels. 
The single channel separation model is trained on the same dataset. This model uses a power spectrum as input and has four BLSTM layers, followed by two mask generation layers. 
The separation processing is applied to each input channel independently, resulting in $2 \times C$ TF masks. 
Because the orders of the separated sources may be inconsistent between different channels, 
cross-channel permutation alignment is performed. We use the separated signals from the first microphone as a reference permutation. For each of the other channels, the output signal order is aligned with that of the first channel by finding the permutation that minimizes the mean squared error between the two magnitude-normalized separation results. 
For each source, the TF masks that yield the highest posterior SNR is picked to perform MVDR beamforming. 


Based on the first baseline system, a second baseline system is built using a multi-channel feature, called a relational feature, in addition to the power spectrum as input to a separation model. 
The relational feature is a weighted combination of other channels' spectra, where the weights are positively correlated with their degrees of similarity to the current channel. The specific formula used in our experiment is given by 
\begin{equation}
\mathbf{Y}_{i} = \sum_{j=0, j\neq i}^{C-1} \mathbf{X}_{j}\cdot \text{softmax} \left(\frac{1}{\parallel \mathbf{X}_{i} - \mathbf{X}_{j}\parallel}\right), 
\end{equation}
where $\mathbf{X}_i$ are the normalized spectrum of the $i$th channel. For each channel $i$, the relational feature $\mathbf{Y}_i$ is concatenated with the spectral feature $\mathbf{X}_i$ to form an input to the separation model. 
The model configuration of the second baseline system is the same as that of the first one.

In addition, to further investigate the efficacy of proposed interleave architecture, we developed a system which includes the same number of self-attention and BLSTM layers, with no interleaving connection, i.e. we stack all the self-attention layers at the bottom, and then add all the BLSTM layers above. Finally, the mean pooling is used to combine all the channels of the signals and used for separation. We call this system "non-interleave system". 


\begin{table}[t]
\centering
\caption{\%WERs of different systems.}
\label{tab:my_label}
\begin{tabular}{lcc}
\hline\hline
System & TF Masking & MVDR \\ \hline
Mixture & \multicolumn{2}{c}{96.93} \\ \hline
Multi-stream & 28.66 & 24.18 \\ 
~~~~+ Relational features & 26.90 & 21.63 \\ \hline
Proposed & \textbf{17.75} & \textbf{16.11} \\ 
~~~~ w/o interleaving structure & 24.39 & 20.89 \\ \hline
\end{tabular}
%
\vspace{2.5em}
\caption{\%WERs with different channel selection schemes.}
\label{tab:table_channel}
\begin{tabular}{lccc}
\hline \hline
System & Oracle & Max-SNR & Random \\ \hline
Multi-stream & 21.68 & 24.18 & 30.66 \\
~~~~ + Relational features  & \textbf{15.06} & 21.63 & 24.23 \\
Proposed & 16.71 & \textbf{16.11} & \textbf{17.85} \\ \hline
\end{tabular}
\end{table}

\subsection{Evaluation}
\label{ssec:eval}

Word error rate (WER) is used as a separation performance metric. 
Our ASR system is based on a conventional hybrid system consisting of a latency-controlled BLSTM acoustic  model (AM) \cite{Xue17} and a weighted finite state transducer decoder.
Our AM is trained on 33K hours of in-house audio data, including close-talking, distant-microphone, and artificially noise-corrupted speech. Decoding is performed with a 5-gram language model (LM).

\subsection{Results}
\label{ssec:results}

Table~\ref{tab:my_label} shows the WERs of different systems. 
All separation systems improved the WERs compared with directly recognizing the original mixed signals.
The mask-based MVDR enhancement scheme outperforms the TF masking as observed in some previous studies. 
The proposed method significantly outperformed the two baseline systems irrespective of the enhancement schemes.
The second baseline system using the relational features
leverages cross-channel information before and after the neural network processing. 
The fact that the proposed method outperformed it indicates 
the effectiveness of modeling the multi-channel information within the network. 
The results also clearly show the benefit of alternately 
exploring the spatial and temporal features by stacking
the spatio-temporal processing blocks. 




\begin{figure}[t!]
  \centering
  \includegraphics[width=0.45\textwidth]{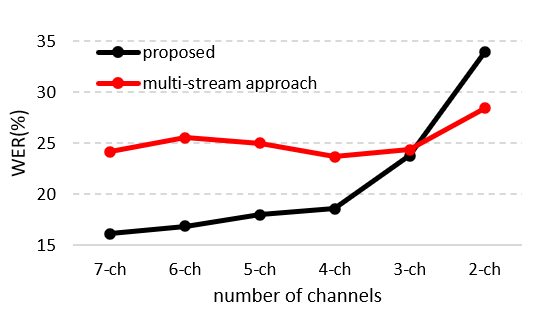}
  \caption{\%WERs for different channel numbers.}
  \label{fig:results_varied_channel}
  \vspace{1em}
\end{figure}

Table~\ref{tab:table_channel} shows the WERs obtained with different reference channel selection schemes, revealing the impact that the reference channel selection accuracy has on the MVDR performance. 
The oracle channel selection method picks the channel that gives the highest signal-to-distortion ratio (SDR) between a TF-mask output signal and a reference clean signal\footnote{Note that the oracle channel selection scheme used TF-masked signals for SDR computation while our automatic channel selection scheme (Max-SNR) were based on MVDR outputs, which could account for the proposed method not benefiting from the oracle channel selection in Table \ref{tab:table_channel}.}.
We can see that 
the proposed method was much more insensitive to  channel selection errors while 
the performance of the multi-stream baseline systems severely 
deteriorated 
when non-oracle channel selection methods were used. 

Figure \ref{fig:results_varied_channel} shows 
the WERs for different numbers of input channels 
obtained with the multi-stream and proposed systems. 
In each evaluation setting, it was guaranteed that there was a relatively close microphone on the table in front of each loudspeaker. For the proposed method, using more microphones consistently improved the WER. However, the performance of the multi-stream system did not change very much depending on 
the microphone number (the best performance was achieved with four microphones). With the multi-stream processing, 
although the use of more microphones should contribute to MVDR accuracy improvement, the gain was offset by the increased difficulty of channel selection. 
In contrast, as the proposed method is far less sensitive to the channel selection accuracy as shown in Table \ref{tab:table_channel},
it always benefited from the increased number of input channels. 
The performance degradation for the two-channel case is one of the areas that need to be addresses in the future while it may be alleviated by including two-channel examples in the training data.

\section{Conclusion}
\label{sec:conclusion}

We presented a neural network based speech separation algorithm using spatially distributed microphones.
A novel spatio-temporal processing block was proposed to take advantage of multi-channel information in a way that is independent of the number and permutation of the microphones. 
Our proposed block consists of a self-attention-based inter-channel processing layer and a temporal processing layer. 
Multiple spatio-temporal processing blocks are stacked, followed by a mean pooling layer for global information fusion. 
Significant WER improvement was achieved compared with the baseline systems. 

\section{Acknowledgement}

We thank Tianyan Zhou and Xiaofei Wang for their contribution to our data collection effort.

\bibliographystyle{IEEEtran}

\bibliography{strings,refs}

\begin{thebibliography}{10}
\providecommand{\url}[1]{#1}
\csname url@samestyle\endcsname
\providecommand{\newblock}{\relax}
\providecommand{\bibinfo}[2]{#2}
\providecommand{\BIBentrySTDinterwordspacing}{\spaceskip=0pt\relax}
\providecommand{\BIBentryALTinterwordstretchfactor}{4}
\providecommand{\BIBentryALTinterwordspacing}{\spaceskip=\fontdimen2\font plus
\BIBentryALTinterwordstretchfactor\fontdimen3\font minus
  \fontdimen4\font\relax}
\providecommand{\BIBforeignlanguage}[2]{{%
\expandafter\ifx\csname l@#1\endcsname\relax
\typeout{** WARNING: IEEEtran.bst: No hyphenation pattern has been}%
\typeout{** loaded for the language `#1'. Using the pattern for}%
\typeout{** the default language instead.}%
\else
\language=\csname l@#1\endcsname
\fi
#2}}
\providecommand{\BIBdecl}{\relax}
\BIBdecl

\bibitem{ty_icassp_2018}
T.~Yoshioka, H.~Erdogan, Z.~Chen, and F.~Alleva, ``Multi-microphone neural
  speech separation for far-field multi-talker speech recognition,'' in
  \emph{Proc. IEEE International Conference on Acoustics, Speech and Signal
  Processing (ICASSP)}, 2018, pp. 5739--5743.

\bibitem{Bahmaninezhad2019}
F.~Bahmaninezhad, J.~Wu, R.~Gu, S.-X. Zhang, Y.~Xu, M.~Yu, and D.~Yu, ``A
  comprehensive study of speech separation: Spectrogram vs waveform
  separation,'' in \emph{Proc. Interspeech}, 2019, pp. 4574--4578.

\bibitem{dyu_taslp_2017}
M.~Kolbæk, D.~Yu, Z.~Tan, and J.~Jensen, ``Multitalker speech separation with
  utterance-level permutation invariant training of deep recurrent neural
  networks,'' \emph{IEEE/ACM Transactions on Audio, Speech and Language
  Processing}, vol.~25, no.~10, pp. 1901--1913, October 2017.

\bibitem{6516079}
M.~{Souden}, S.~{Araki}, K.~{Kinoshita}, T.~{Nakatani}, and H.~{Sawada}, ``A
  multichannel {MMSE}-based framework for speech source separation and noise
  reduction,'' \emph{IEEE Transactions on Audio, Speech, and Language
  Processing}, vol.~21, no.~9, pp. 1913--1928, 2013.

\bibitem{HEYMANN2017374}
J.~Heymann, L.~Drude, and R.~Haeb-Umbach, ``A generic neural acoustic
  beamforming architecture for robust multi-channel speech processing,''
  \emph{Computer Speech, Language}, vol.~46, pp. 374--385, 2017.

\bibitem{erdogan_interspeech_2016}
H.~Erdogan, J.~Hershey, S.~Watanabe, M.~Mandel, and J.~Le~Roux, ``Improved
  {MVDR} beamforming using single-channel mask prediction networks,'' in
  \emph{Proc. Interspeech}, 2016, pp. 1981--1985.

\bibitem{ty_interspeech_2018}
T.~Yoshioka, H.~Erdogan, Z.~Chen, X.~Xiao, and F.~Alleva, ``Recognizing
  overlapped speech in meetings: A multichannel separation approach using
  neural networks,'' in \emph{Proc. Interspeech}, 2018, pp. 3038--3042.

\bibitem{PrincetonASRU2019}
T.~Yoshioka, I.~Abramovski, C.~Aksoylar, Z.~Chen, M.~David, D.~Dimitriadis,
  Y.~Gong, I.~Gurvich, X.~Huang, Y.~Huang, A.~Hurvitz, L.~Jiang, S.~Koubi,
  E.~Krupka, I.~Leichter, C.~Liu, P.~Parthasarathy, A.~Vinnikov, L.~Wu,
  X.~Xiao, W.~Xiong, H.~Wang, Z.~Wang, J.~Zhang, Y.~Zhao, and T.~Zhou,
  ``Advances in online audio-visual meeting transcription,'' in \emph{Proc.
  IEEE Automatic Speech Recognition and Understanding Workshop (ASRU)}, 2019.

\bibitem{zliu_iwaenc_2008}
Z.~Liu, ``Sound source seperation with distributed microphone arrays in the
  presence of clocks synchronization errors,'' in \emph{Proc. International
  Workshop for Acoustic Echo and Noise Control (IWAENC)}, 2008, p. 14–17.

\bibitem{Araki18-ICASSP}
S.~{Araki}, N.~{Ono}, K.~{Kinoshita}, and M.~{Delcroix}, ``Meeting recognition
  with asynchronous distributed microphone array using block-wise refinement of
  mask-based {MVDR} beamformer,'' in \emph{Proc. IEEE International Conference
  on Acoustics, Speech and Signal Processing (ICASSP)}, April 2018, pp.
  5694--5698.

\bibitem{Yoshioka19c}
\BIBentryALTinterwordspacing
T.~Yoshioka, Z.~Chen, D.~Dimitriadis, W.~Hinthorn, X.~Huang, A.~Stolcke, and
  M.~Zeng, ``Meeting transcription using virtual microphone arrays,''
  \emph{CoRR}, vol. abs/1905.02545, 2019. [Online]. Available:
  \url{http://arxiv.org/abs/1905.02545}
\BIBentrySTDinterwordspacing

\bibitem{yiluo_icassp_2020}
Y.~Luo, Z.~Chen, N.~Mesgarani, and T.~Yoshioka, ``End-to-end microphone
  permutation and number invariant multi-channel speech separation,'' in
  \emph{Proc. IEEE International Conference on Acoustics, Speech and Signal
  Processing (ICASSP)}, 2020.

\bibitem{9003849}
Y.~{Luo}, C.~{Han}, N.~{Mesgarani}, E.~{Ceolini}, and S.~{Liu}, ``{FaSNet}:
  Low-latency adaptive beamforming for multi-microphone audio processing,'' in
  \emph{Proc. IEEE Automatic Speech Recognition and Understanding Workshop
  (ASRU)}, 2019, pp. 260--267.

\bibitem{transformer_2017}
A.~Vaswani, N.~Shazeer, N.~Parmar, J.~Uszkoreit, L.~Jones, A.~N. Gomez,
  L.~Kaiser, and I.~Polosukhin, ``Attention is all you need,'' in \emph{Proc.
  NeurIPS}, 2017, p. 1–11.

\bibitem{Sak14}
H.~Sak, A.~Senior, and F.~Beaufays, ``Long short-term memory recurrent neural
  network architectures for large scale acoustic modeling,'' in \emph{Proc.
  Interspeech}, 2014, pp. 338--342.

\bibitem{mask_2013}
A.~Narayanan and D.~Wang, ``Ideal ratio mask estimation using deep neural
  networks for robust speech recognition,'' in \emph{Proc. IEEE International
  Conference on Acoustics, Speech and Signal Processing (ICASSP)}, 2013, pp.
  7092--7096.

\bibitem{roux2018sdr}
J.~L. Roux, S.~Wisdom, H.~Erdogan, and J.~R. Hershey, ``{SDR} - half-baked or
  well done?'' in \emph{Proc. IEEE International Conference on Acoustics,
  Speech and Signal Processing (ICASSP)}, 2019.

\bibitem{7404828}
T.~{Yoshioka}, N.~{Ito}, M.~{Delcroix}, A.~{Ogawa}, K.~{Kinoshita},
  M.~{Fujimoto}, C.~{Yu}, W.~J. {Fabian}, M.~{Espi}, T.~{Higuchi}, S.~{Araki},
  and T.~{Nakatani}, ``The {NTT CHiME-3 system}: Advances in speech enhancement
  and recognition for mobile multi-microphone devices,'' in \emph{Proc. IEEE
  Workshop on Automatic Speech Recognition and Understanding (ASRU)}, 2015, pp.
  436--443.

\bibitem{panayotov2015librispeech}
V.~Panayotov, G.~Chen, D.~Povey, and S.~Khudanpur, ``Librispeech: an {ASR}
  corpus based on public domain audio books,'' in \emph{Proc. IEEE
  International Conference on Acoustics, Speech and Signal Processing
  (ICASSP)}, 2015, pp. 5206--5210.

\bibitem{image_method_1979}
J.~B. Allen and D.~A. Berkley, ``Image method for efficiently simulating
  small-room acoustics,'' \emph{Journal Acoustic Society of America}, vol.~65,
  no.~4, pp. 943--950, April 1979.

\bibitem{gpu_rir}
D.~Diaz-Guerra, A.~Miguel, and J.~R. Beltran, ``gpu{RIR}: A python library for
  room impulse response simulation with {GPU} acceleration,'' in \emph{arXiv},
  2018.

\bibitem{chen2020continuous}
Z.~Chen, T.~Yoshioka, L.~Lu, T.~Zhou, Z.~Meng, Y.~Luo, J.~Wu, and J.~Li,
  ``Continuous speech separation: dataset and analysis,'' in \emph{Proc. IEEE
  International Conference on Acoustics, Speech and Signal Processing
  (ICASSP)}, 2020.

\bibitem{Xue17}
S.~{Xue} and Z.~{Yan}, ``Improving latency-controlled {BLSTM} acoustic models
  for online speech recognition,'' in \emph{Proc. IEEE International Conference
  on Acoustics, Speech and Signal Processing (ICASSP)}, 2017, pp. 5340--5344.

\end{thebibliography}

\end{document}